# Doppler Shift or Another Mass of a Nucleon?

A.V.Kopylov (Kopylov@al20.inr.troitsk.ru)
Institute for Nuclear Research of RAS, Moscow, Russia,
117312, Prospect of 60[th] Anniversary of October Revolution 7a

**Abstract.**

The comparison of magnetic moments of neutron and proton reveals three possible stationary states of a nucleon: s-, p- and o-nucleon with different configurations and presumably different masses. Two of these nucleons may exist as the small impurities to the representative one. If so the frequency 1420405751.767 Hz of the hydrogen maser should be accompanied by the lines of weak intensity with a frequency shift proportional to the mass shift of a nucleon state. In radio astronomy the manifestation of this can be the presence of the corresponding "Doppler shifted" lines in the microwave spectrum of hydrogen.

It was shown in Ref.1 that if to take the model of Ida and Kobayashi [2] and to compare the magnetic moments of neutron and proton, taking into account that these particles are different members of one isospin multiplet, then this comparison reveals three possible states of a nucleon: one with a single quark at the center of nucleon and a scalar diquark composed of like quarks, moving around (lets call this state s-nucleon), second - with a diquark ud in $^3S_1$ state and a single quark with the spin up, moving around (p-nucleon) and third - the same as the second one only the spin of a single quark is down (o-nucleon). This result is obtained by using no free parameters, the only supposition (apart from the model of Ida and Kobayashi) used is that the magnetic moment of a nucleon can be expressed as

$$\mu = \mu_N \cdot \Sigma e_i g_i \mathbf{J}_i \qquad (1)$$

here $e_i$, $g_i$ and $\mathbf{J}_i$ are the charge, factor of Lande and moment of a constituent quark, $\mu_N$ – is a nuclear magneton. This expression assumes that the only representative mass here is a mass of a nucleon. This supposition is quite natural taking into consideration the strong forces between quarks. For a non disturbed nucleon (the case of $q^2=0$) the quarks inside can not move free of the rest of a nucleon mass.

Let's take into account that diquark can exist as a pair of like quarks and as a pair of different quarks and write the corresponding system of two equations for both cases taking into consideration that neutron and proton are members of one isospin multiplet:

Case I:  $n = u + dd$   $p = d + uu$

$$\begin{cases} 2/3\, g_1 J_1 - 2/3\, g_2 J_2 = -1.913 \\ -1/3\, g_1 J_1 + 4/3\, g_2 J_2 = 2.793 \end{cases}$$

From here it follows that: $g_1 J_1 = -1.033$ and $g_2 J_2 = 1.836$
The fact that $g_1 J_1$ turned out to be very close to 1 means that here the single quark acts as elementary particle placed at the center of nucleon, some extra of 0.033 may be assumed here as a glueball correction similar to the Schwinger correction to magnetic moment of electron. Presumably this is the consequence of the quark recoil whenever a virtual quark-antiquark pair is emitted [3]. By this recoil the quark acquires some electromagnetic formfactor which leads to this correction 0.033. The fact that $g_2 J_2$ turned out to be less than 2 means that here diquarks are scalars with spin zero, i.e. that spins of quarks in a diquark are antiparallel what agrees with the Pauli principle.

Case II. The similar equations can be written for the case when diquark **ud** is in a state $^3S_1$, only here $g_1 J_1$ is the value assigned to the single quarks, constituents of a diquark **ud**:

$n = ud + d$   $p = ud + u$

$$\begin{cases} 1/3\, g_1 J_1 - 1/3\, g_2 J_2 = -1.913 \\ 1/3\, g_1 J_1 + 2/3\, g_2 J_2 = 2.793 \end{cases}$$

To satisfy this system one should take: $g_1 J_1 = -1.033$ and $g_2 J_2 = 4.7$. The fact that $g_1 J_1$ turned out to be close to 1 (with the same glueball correction) agrees perfectly with the assumption $^3S_1$. In fact, we have two more states in this case: one with a spin of a single quark up, and another one - with a spin down. But for both these states there is no integer L which would correspond to the value $g_2 J_2 = 4.7$. Because for a given $g_2 J_2 = 4.7$ we will get different L for different orientation of spin of a single quark, this is a clear indication on spin-dependent forces between quark and diquark. The spin-dependence of the forces follows also from the fact that **ud** diquark can be constituent of nucleon (and occupy the place at the center of nucleon) only when spins of both quarks are parallel. Good argument in favor that the model presented here is realistic is that one finds agreement with the measured magnetic moments of both nucleons if quarks at the center of nucleon act as elementary particles, this is observed here for all three states. Qualitatively this conforms the parton model of the nucleon and $x$-scaling behavior in deep-inelastic lepton scattering first observed over twenty years ago with electrons[4]. Figure 1 of Ref.1 illustrates these states.

Now, the question is: what are these states and what will be the evolution equations for these states

$$idC_j/dt = H_{jk}C_k, \qquad (2)$$

in other words, what Hamiltonian we will have here? This question was discussed in Ref.5.

The remarkable thing is that in all three states we have something common: the single quark is always in the center of nucleon and its spin is always oriented antiparallel to the spin of a nucleon. It looks like we have a couple of like quarks with different motion in a strong field of a single quark at the center of nucleon which does not change its orientation. But this picture looks familiar. We have something of this kind when we put a hydrogen atom in a strong magnetic field. It is well known that one has a hyperfine splitting in this case: the state S=1 is splitting to three states with m=+1, 0 and −1 and these are stationary states as well as another state with S=0. It is important that while at low field the stationary states S=1, m=0 and S=0 are the mixture of states |+−> and |−+> :

$$S=1, m=0: \qquad \frac{1}{\sqrt{2}}(|+-> + |-+>)$$

$$S=0: \qquad \frac{1}{\sqrt{2}}(|+-> - |-+>)$$

(here the first sign denotes the orientation of spin of electron, the second − of proton), at high field they are not mixed, so that the first state (S=1, m=0) is a pure |+−> state while the second (S=0) is a pure |−+> state. This similarity looks intriguing, at least what concerns the p- and o-nucleons. It can explain why in contrast with a ground state of a hydrogen atom where we have four stationary states, here we have only two. The diquark at the center of nucleon is always in a state $^3S_1$ with the spins of both quarks antiparallel to a spin of a nucleon. It means that in this case we don't have the states |++> and |—+>, (here the first sign denotes the orientation of spin of peripheral quark, the second – of central quark of a pair of like quarks). And in addition to these two states we have one more state (s-nucleon) with a peripheral pair of like quarks with antiparallel spins (see fig.1 in Ref.1).

So if to follow the analogy with a hydrogen atom in a strong magnetic field the p- and o-nucleon are the stationary states of a nucleon if to assume that there is no mixing with s-nucleon. The Hamiltonium in this case should be diagonal. This case conforms the straightforward comparison of magnetic moments of proton and neutron by which these p-, o- and s-states were revealed.

Now, what one can say about the possible masses of these nucleons? First-of-all we should keep in mind that the current accuracy of a nucleon mass measurement is 0.28 keV [6], which is far below the energy of the interaction between magnetic moments of quarks in a nucleon. So if another two states do exist they should be only small impurities to the representative nucleon. By very crude estimates, if to take into account only the electromagnetic interaction of magnetic moments of quarks inside of a nucleon, the difference of masses of p- and o-nucleons should be about 100 keV. This value gives only a scale of possible mass difference. From here it follows, that the concentration of the impurities are allowed on the level less than 1%. It is very interesting to find an experimental way to look for these new nucleons. Luckily the mass of proton is directly connected with the hyperfine splitting between the spin-0 and spin-1 levels in the hydrogen ground state which is one of the most accurately measured quantities in physics. The frequency of the hydrogen maser is found to be[7]:

$$f_H = 1\ 420\ 405\ 751.766\ 7 \pm 0.001\ Hz$$

This frequency was calculated with a good accuracy[8] also but the accuracy of the calculation is about six orders of magnitude below the one of measurements. According to calculation:

$$\Delta E(QED) = E_F \{1 + 3(Z\alpha)^2/2 + \ldots\}$$

where $E_F$ is the Fermi splitting

$$E_F = \frac{8}{3}\alpha^4 c^2\ (1+k)\frac{(m_e m_p)^2}{(m_e + m_p)^3}$$

here $\alpha$ is the fine-structure constant, $m_e$ and $m_p$ are the electron and proton masses, k is the proton's anomalous moment coefficient. As one can see from these expressions for small shift of the mass the frequency shift is proportional (with the sign minus) to the shift of the mass. So by scanning the microwave spectrum of hydrogen is possible to find the weak lines which correspond to another configurations of a nucleon. It looks reasonable to scan a range about ±1 MHz from the main line. In radio astronomy these lines (if they do exist) may look like the Doppler shifted lines in the microwave spectrum of hydrogen.

I am grateful to V.A.Kuzmin for the stimulating discussions.